\documentclass[preprint2]{aastex}

\newcommand{\muas}{$\mu$as}

\newcommand{\masa}{mas$\cdot$yr$^{-1}$}
\newcommand{\ms}{m\,s$^{-1}$}
\newcommand{\msun}{M$_{\odot}$}
\newcommand{\mjup}{M$_{\mbox \footnotesize J}$}
\newcommand{\msini}{M$\cdot$\,sin\,$i$}
\newcommand{\bt}{\hbox{$B_{\mbox \footnotesize T}$}}
\newcommand{\vt}{\hbox{$V_{\mbox \footnotesize T}$}}

\newcommand{\itsim}{\hbox{\it SIM}}
\newcommand{\hipparcos}{\hbox{\it Hipparcos}}

\shortauthors{Frink et al.}
\shorttitle{A Strategy for Identifying {\it SIM} Grid Stars}

\begin{document}

\title{A Strategy for Identifying the Grid Stars for the\\Space Interferometry Mission (\itsim)}

\author{Sabine Frink and Andreas Quirrenbach}
\affil{University of California San Diego, La Jolla, CA 92093, USA}
\email{sabine@ucsd.edu, aquirrenbach@ucsd.edu}

\author{Debra Fischer}
\affil{University of California at Berkeley, Berkeley, CA 94720, USA}
\email{fischer@serpens.berkeley.edu}

\author{Siegfried R\"oser}
\affil{Astronomisches Rechen-Institut Heidelberg, 69120 Heidelberg, Germany}
\email{s19@ix.urz.uni-heidelberg.de}

\and

\author{Elena Schilbach}
\affil{Astrophysikalisches Institut Potsdam, 14482 Potsdam, Germany}
\email{eschilbach@aip.de}

\begin{abstract}
We present a strategy to identify several thousand stars that are
astrometrically stable at the micro-arcsecond level for use in the
\itsim\ (Space Interferometry Mission) astrometric grid.
The requirements on the grid stars make this a rather challenging task.
Taking a variety of considerations into account we argue for K~giants as the
best type of stars for the grid, mainly because they can be
located at much larger distances than any other type of star due to their
intrinsic brightness. We show that it is possible to identify suitable
candidate grid K~giants from existing astrometric catalogs.
However, double stars have to be eliminated from these candidate grid samples, 
since they generally produce much larger astrometric jitter than
tolerable for the grid. The most efficient way to achieve this is probably by
means of a radial velocity survey. To demonstrate the feasibility of this
approach, we repeatedly measured the radial velocities for a 
pre-selected sample of 86~nearby \hipparcos\ K~giants with precisions of 5--8\,\ms.
The distribution of the intrinsic radial velocity variations for the bona-fide
single K~giants shows a maximum around 20\,\ms, 
which is small enough not to severely affect the identification of 
stellar companions around other K~giants. We use the results of our observations as
input parameters for Monte-Carlo simulations on the possible design of a radial
velocity survey of all grid stars. Our favored scenario would result in a grid
which consists to 68\% of true single stars and to 32\% of double or multiple stars 
with periods mostly larger than 200~years, but only 3.6\% of all grid stars
would display astrometric jitter larger than 1\,\muas. 
This contamination level is probably tolerable.
\end{abstract}


\section{Introduction}

The Space Interferometry Mission (\itsim, see e.\,g.\ \citealt{simbook}),
which is currently scheduled for launch around 2008/2009 and 
one of the next major steps in 
NASA's Origins Program \citep{origins}, 
is designed to perform astrometry at the 4\,\muas\
level in its wide-angle mode and at the 1\,\muas\ level in its narrow-angle mode.
The instrument is a Michelson Interferometer with a 10\,m baseline, operating
in the visible. 

The detection of extrasolar planets via the astrometric wobble they induce in
their parent stars is one of the main scientific goals of the \itsim\ mission.
The astrometric accuracy at the end of a 5~year mission will allow for the
detection of Jupiter-like planets at distances up to $\approx$\,1\,kpc
(the astrometric signature of Jupiter at 1\,kpc is 10\,\muas) and of planets of
a few Earth masses around the most nearby stars.

The unprecedented accuracy of \itsim, which is two to three orders of magnitude better
than what was achieved with the recent \hipparcos\ astrometric satellite 
\citep{esa}, strongly depends on the astrometric stability of its
grid objects. The grid consists of several thousand stars distributed uniformly
over the sky, with the exact number depending on the final optical design of
the \itsim\ instrument. Dedicated grid campaigns will be carried out
several times a year, with a total of about 110~visits for each star.
The grid then serves as a reference frame to which the individual science
observations can be tied. In addition to enabling wide-angle astrometry
the grid also will lead to improvements in the International
Celestial Reference Frame, ICRF
(see e.g.\ \citealt{gaume} or the review by \citealt{johnston}).

A few radio-loud quasars will be included in the grid to establish a
direct link between the optical reference frame defined by the \itsim\
astrometric grid and the radio-based ICRF. The inclusion of a few quasars in
the grid also helps to eliminate possible spurious rotation of the \itsim\
reference frame. 

Finding well-suited grid stars is by no means an easy task. Improving any
measurement accuracy by several orders of magnitude always implies the chance
(and risk!) of finding something new and unexpected. The aim of this study is
to present a strategy that can minimize any such unwanted surprises for the \itsim\
astrometric grid stars.

The importance of a low contamination level of the grid sample with
astrometrically unstable stars is illustrated in Table~\ref{grid}, where we
calculated the percentages of usable \itsim\ observations for
various contamination levels from basic statistics.
As a minimum four good grid stars are required
for every science observation to solve for the baseline, but some redundancy
would be highly desirable. 
For an assumed contamination level of 5\% of the grid with unusable stars 
and six scheduled grid star observations statistically 99.8\% of all science
observations could be used.
However, with a grid star failure rate of 20\%
the fraction of usable science observations will decrease to only 90\%, wasting
valuable mission time. In order to
increase the success rate to 99\%, eight instead of six grid stars would have
to be observed along with every science observation. At least for bright
science targets this would add significant overhead compared to the
integration time spent for the actual target and would thus significantly
lower the overall number of stars that could be observed with \itsim.
Instead, we propose an intense ground campaign to eliminate astrometrically 
unstable stars from the grid. 

This paper is organized as follows. After arguing for K giants as the
type of stars which is best suited for use with the \itsim\ grid in Section~2, we
define a nearby proxy sample of \hipparcos\ K~giants and present the first
results of a precise radial velocity study in Section~3. In Section~4 we outline a
strategy to identify several thousand well-suited grid stars within the next
few years. Finally we discuss our results in Section~5 and 
provide a short summary in Section~6. 

\begin{deluxetable}{c@{$\qquad$}cccc}
\tabletypesize{\footnotesize}
\tablewidth{210pt}
%\tablewidth{240pt}
\tablecolumns{5}
\tablecaption{\label{grid} Percentages of usable \itsim\ observations}
\tablehead{
no.\ of grid stars & \multicolumn{4}{c}{contamination level} \\
per science obs.\ & \multicolumn{1}{l}{5\%\phn\phn\phn\phn} & \multicolumn{1}{l}{10\%} & 20\% & 50\% 
}
\startdata
4  &  81\phd\phn\phn\phn & 66\phd\phn\phn & 41 &\phn6 \\
5  &  97.7\phn\phn       & 92\phd\phn\phn & 74 &   19 \\
6  &  99.8\phn\phn       & 98\phd\phn\phn & 90 &   34 \\
7  &  99.98\phn          & 99.7\phn       & 97 &   50 \\
8  &  99.999             & 99.96          & 99 &   64 \\
\enddata
\end{deluxetable}

\section{Which Types of Stars would be Good Grid Stars?}

The concept of the \itsim\ Astrometric Grid is described e.g.\ in the \itsim\ Book
\citep{simbook}, \citet{boden99} or the 
\anchor{http://sim.jpl.nasa.gov/}{\itsim\ website\footnote{\url{http://sim.jpl.nasa.gov/}}}.
In addition to the dedicated grid campaigns a minimum of four grid stars will
be observed along with each science target to solve for the spacecraft
orientation. Depending on the size of \itsim's Field of Regard (FOR) this gives
a minimum number of grid stars required to cover the sky. However, a larger
number of stars per tile (corresponding in size to the FOR) is highly
desirable, for redundancy as well as for attrition reasons among grid
stars. The current grid star design anticipates twelve stars for each tile and
a FOR of either 15 or 20~degrees, resulting in a total of approximately 3000 or
1500~grid stars, respectively. 

A larger number of grid stars has the advantage that the separations between
science targets and grid stars are smaller, which increases the astrometric
accuracy. Furthermore, if it turns out during the mission that observations in 
a specific tile are compromised 
because not enough grid stars fulfill the requirements on
astrometric stability, it does not have such a high impact as in the case with
only half as many tiles and grid stars. On the other hand, however, a
larger number of qualifying stars is more difficult to find and requires a
larger amount of observing resources from the ground, before the mission, as well
as longer \itsim\ grid campaigns.

In any case, the grid stars have to be distributed uniformly over the sky.
This requirement, among other shortcomings like the difficulty to measure
precise radial velocities for fast rotating stars, precludes O or B
stars from being grid stars, because these are concentrated in the spiral arms
of the galactic disk.
Furthermore, the grid stars should in general not be fainter than about
12\,mag, because otherwise the fraction of \itsim\ observing time dedicated to
the grid instead of actual science observations would become too large.

The most important and most challenging requirement however is the astrometric
stability of the photocenter of the grid objects to within a few
microarcseconds. \\Most double\footnote{speaking of double or binary stars in this
paper we always mean to include triple, quadruple and all other multiple stars}
stars are therefore unacceptable as grid stars,
because the orbital motion imposes too many uncertainties in the position and
velocity model for a grid star; the residual non-modeled uncertainties in the
positions of the stars, the so-called astrometric jitter, would be much larger
than a few microarcseconds.
Even planets can impose problems, but since the
astrometric signature scales with the inverse of the distance one can get rid
of part of the problem by choosing rather distant grid stars. 
Note that parallactic and linear motions of the grid stars do not impose any
problems since parallax and proper motions can be determined from the \itsim\ 
observations with the required precision and need not be known before
the mission.

Since the
apparent magnitude for grid stars is limited this calls for intrinsically
bright stars like K~giants. In terms of brightness M giants would work even better,
but most of them are highly variable and therefore not very well suited as grid
stars. A 12\,mag K~giant would typically be located at about 2\,kpc, assuming an
absolute magnitude $M_V$ of +0.5\,mag for solar metallicity. 
Metal-weak halo K~giants are up to 2~mag brighter, which places them 
at distances up to 5\,kpc.
The astrometric signature of a Jupiter-like companion, assuming one solar mass for 
the parental star, would be 5\,\muas\ peak-to-peak at 2\,kpc and 2\,\muas\ 
peak-to-peak at 5\,kpc.

This level is still at the higher end of what would be acceptable for
\itsim. While a reasonable strategy exists on how to clean a candidate grid
star sample from unwanted multiple systems, there is no way to assure the absence
of planets for such a large sample of stars with the instruments on hand at
present. Brown Dwarfs however are not a big concern, see Section~\ref{bddsec}.

K~giants are known to be photospherically active. This leads to intrinsically
variable radial velocities, which could complicate the identification of
binaries among a sample of K~giants. Sections~3 and 4 deal with this problem
and show that this is not a severe complication. Another related effect is the
photocenter shift induced by starspots. However, most starspots are not large
and cool enough to produce measurable photocenter shifts at distances of
several kpc.

There is no reasonable alternative to using K~giants as grid objects.
Bright G~dwarfs, which had also been suggested as grid stars in earlier phases
of the project, would be advantageous in terms of already available information 
as well as brightness since that they could be easily observed at moderately sized 
telescopes. Most stellar companions could probably be identified using already
available data, and one might even be able to exclude giant Jupiters as
companions by selecting the rejects of current planet searches.
However, given the magnitude limit of 12\,mag for grid stars, G~dwarfs would be
located at distances not larger than 300\,pc.
The astrometric signature of a Jupiter at that distance would be
more than 30\,\muas, clearly much larger than acceptable. For brighter and thus
closer G~dwarfs even Saturn-mass planets would generate detectable 
astrometric signals. Furthermore, we
already know now that planetary companions to solar-like stars are numerous;
these are the stars monitored by the radial velocity planet search programs 
\citep[see e.g.][]{mayor99,bio99}. 

\begin{figure}[t]
\plotone{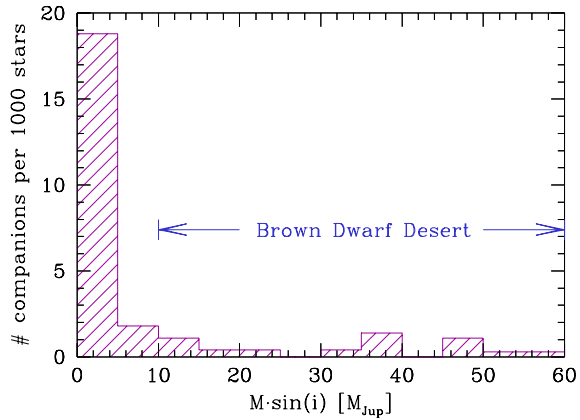}
\figcaption{\label{companions} 
Histogram of substellar companions that have been found
by radial velocity searches, normalized by the number of stars
surveyed. Included are the results from planet searches using the Hamilton
Spectrograph at Lick Observatory, HIRES at Keck, the Coravel, Elodie and
Coralie spectrographs at Observatoire de Haute-Provence and at La Silla,
CES at La Silla, the AFOE spectrograph at Whipple Observatory, and at the
Smithsonian Astrophysical Observatory. The lack of systems in the Brown Dwarf
mass range is clearly evident, especially since the radial velocity surveys are
more sensitive to higher masses and the diagram is fairly incomplete
towards lower masses. The elimination of the sin\,$i$ uncertainty is not likely
to change the appearance of this diagram significantly.}
\end{figure}

\subsection{Brown Dwarf Desert}
\label{bddsec}

We have plotted the number of companions found in the planet and Brown Dwarf
mass ranges normalized by the number of stars surveyed for all radial velocity
planet searches that have found planets or Brown Dwarf companions so far 
(Fig.~\ref{companions}). Note that
the surveys are still very incomplete in the planet mass range; planet
identifications just continue to drop out of these surveys with time. On the
other hand, however, the surveys are very sensitive to companions in the Brown
Dwarf mass range, yet very few --- if any after elimination of the sin\,$i$
uncertainty --- have been found. It is unlikely that a significant number of
planetary companions with \msini\ below 10\,\mjup\ will be shifted to the Brown
Dwarf regime, since low values for sin\,$i$ are really rare.

\begin{figure}[t]
\plotone{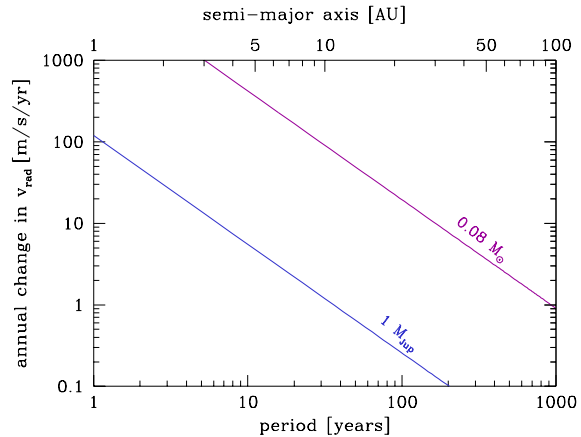}
\figcaption{\label{bdd} Approximate trends that would be observable in the
radial velocities of solar-type stars of 1\,\msun\ if they had a companion
with masses of 1\,\mjup\ and 0.08\,\msun, respectively. Inclinations of 
90\arcdeg\ and circular orbits have been assumed in calculating the annual radial
velocity changes. The semi-major axis scale at the top corresponds to the case
of a 1\,\mjup\ companion; for the companion of 0.08\,\msun\ it would have to be
increased by about 5\% to match the period scale at the bottom. Current radial
velocity surveys would probably be able to detect Brown Dwarfs up to
separations of 30--50~AU, corresponding to periods up to about 200~years.
}
\end{figure}

However, radial velocity surveys are much more sensitive towards smaller
separations, and very little is known so far about Jupiter mass planets beyond
about 5~AU. Brown Dwarfs would still produce a detectable radial
velocity signal at separations up to about 30--50\,AU, or periods of about
200~years (see Fig.~\ref{bdd}). Beyond, the annual changes in the radial
velocities would be too small to be recognized over timescales of a few years.
Astrometric detections probe a different parameter space since the astrometric 
signal is larger for larger separations, although for astrometric detections
the large involved periods are just as limiting as they are for the radial velocity
detections. Other techniques like direct imaging or the 2MASS Survey have
revealed only very few Brown Dwarf companions so far 
\citep{oppenheimer98,rebolo98,burgasser00}.
\citet{halbwachs00} derived astrometric
orbits for eleven spectroscopic binaries with Brown Dwarf Candidates from
\hipparcos\ data and concluded that there is a real minimum in the mass
distribution of companions to solar-type stars.

Thus it seems to be justified to speak of a Brown Dwarf Desert
\citep{mb2000} --- while numerous isolated Brown Dwarfs
exist, very few seem to be companions to solar-like stars. Since these
solar-type stars are the precursors of K~giants, this finding should equally
well hold true for them. So if we are able to sieve a sample of grid star
candidates for the stellar companions, we can take advantage of the Brown
Dwarf Desert. The only companions that are left in the sample are planetary
companions, and their disturbing effects can be minimized by placing the stars
at the largest possible distances.

\subsection{How to Identify Suitable K~Giants}

The most precise and comprehensive astrometric catalogs available today are the 
Hipparcos and Tycho-2 Catalogues \citep{esa,hog}, which makes them a natural
first choice when searching for astrometric reference stars.
The Hipparcos Catalogue contains about 120\,000~stars with an
accuracy of $\approx$1\,mas for positions and parallaxes and $\approx$1\,\masa\
for proper motions. It also provides a wealth of supplementary information
like spectral types, variability or duplicity information, which allows for a
thorough pre-selection of suitable reference stars. 
Comparison of the instantaneous \hipparcos\ proper motion with proper motions
derived from larger epoch differences, as provided e.g.\ by ACT \citep{act} 
or TRC \citep{trc}, can reveal
astrometric binaries \citep[cf.][]{wielen99}.
The parallaxes can be used to distinguish K~giants from K~dwarfs.

Exploiting all the information available, we are left with a sample of
11\,813~candidate K~giant grid stars out of a total of 29\,466~K~giants present
in the Hipparcos Catalogue. The full set of applied selection criteria is given
in the appendix (see also \citealt{frink99}).
The selection criteria work
somewhat better for brighter stars, for which on average more information is
available than for the fainter part of the sample: among the stars brighter
than 6\,mag 87\,\% have been rejected as possible grid stars, while the average
percentage of rejected grid stars is 60\,\% for the whole sample.
However, since we wanted to demonstrate the possibility to find grid stars
which fulfill the requirements of the \itsim\ grid, we concentrated on the very
best grid candidates and chose rather strict selection criteria, 
so that it is likely that also a number of in fact well-suited 
reference stars got rejected. 

\noindent The use of the sample defined in this way is two-fold:

(1) The brighter end of the sample can be used as a proxy sample to study the
    properties of K~giants as grid stars in some detail. 177~\hipparcos\ K~giants
    brighter than 6\,mag were selected as proxies for possible grid stars. 
    We started to monitor the intrinsic radial
    velocity variations in this sample of bona-fide single stars at Lick
    Observatory, and the first results of these observations are presented in
    the next section.

(2) On the other hand, the faint end of the sample contains actually good
    astrometric reference stars at distances as needed for the \itsim\ grid, 
    i.e.\ larger than $\approx$\,1--2\,kpc. 
    We computed photometric distances for these
    stars using the spectral type given in Hipparcos under the assumption of
    solar metallicity. For metal-weak K~giants these distances are a lower
    limit only, whereas for stars with a considerable amount of extinction
    along the line of sight the computed distances will be too large. However,
    the accuracy is good enough to sort out most K~giants which are not
    located at the required minimum distance, and we ended up with a sample of 
    only 45~\hipparcos\ K~giants with computed distances larger than 1\,kpc.
    These could be actually good grid candidates, provided that the radial
    velocities show no variability.

It is obvious that not enough \itsim\ grid stars can be found in the Hipparcos
Catalogue; although the magnitude limit is 12.4\,mag, it is complete only to
7.3--9\,mag, depending on position on the sky. Recently, the Tycho-2 Catalogue
\citep{hog} has been published, which contains
2.5\,million stars with a completeness level of 90\,\% at 11.5\,mag.
It is based on the same observational material as the Tycho-1
Catalogue, supplemented with 144 ground based astrometric
catalogues. Furthermore, a more rigorous reduction has been applied for Tycho-2, 
leading to astrometric
accuracies of 60\,mas for positions and 2.5\,\masa\ for proper motions.
Color information in the Tycho $B_T$ and $V_T$ passbands allows for the
identification of K~stars, while the proper motions can be used to distinguish
between K~giants and K~dwarfs in a statistical sense. Comparison with the
Hipparcos data showed that less than 10\,\% of either K~dwarfs or K~giants
would be misclassified because of extraordinarily small or large proper
motions, respectively. Since there are about 10~times as many K~giants as
K~dwarfs in the magnitude-limited samples considered here, 
this results in a very small contamination fraction of a K~giant sample
with K~dwarfs of about 1\,\%.

Applying a similar set of selection criteria as for the \hipparcos\ stars
(see appendix),
we were able to identify a sample of $\approx$8000~K~giants from Tycho-2 with
computed distances larger than 1.5\,kpc, see also \citet{frink00}. 
This sample of promising \itsim\ grid stars could
be used as a starting point for a radial velocity survey which then has to 
identify the truly single stars among the grid candidates. 

A completely different approach is followed by Majewski, Patterson and co-workers 
\citep{ggss}. They are currently conducting a patchy
all-sky survey at Las Campanas Observatory, aimed at identifying distant Halo
K~giants for the \itsim\ grid. This is accomplished by a photometric
survey with a special set of filters, followed by a spectroscopic verification
of the K~giant nature. Most of the stars are metal-weak K~giants around
12\,mag. 

These large distances make the sample of stars very well suited for the \itsim\
grid; at least in terms of distances and corresponding anticipated astrometric
jitter it is superior to the Tycho-2 sample. However, no variability or
duplicity information whatsoever is available for the Halo K~giant sample, and
it will probably take more observational effort from the ground to clean this
sample from unwanted double and multiple stars.
It would be ideal if it turned out that there was a considerable overlap
between the Halo K~giant and the Tycho-2 sample, combining the advantages of
both stellar samples. 

\section{A Nearby Proxy Sample}
\label{proxy}

\subsection{Observational Results}

\begin{deluxetable}{rccr@{$\qquad$}|@{$\qquad$}rccr@{$\qquad$}|@{$\qquad$}rccr@{}}
\tabletypesize{\footnotesize}
\tablewidth{469.75499pt}
%\tablewidth{515pt}
\tablecolumns{12}
\tablecaption{\label{obstab}Intrinsic Radial Velocity Dispersions for Hipparcos K~giants 
in our proxy sample}
\tablehead{
HIP & &
\multicolumn{1}{c}{$\Delta v_{\rm rad}$} &
\multicolumn{1}{l}{\phn$\sigma_{v_{\rm rad}}$} &
HIP & &
\multicolumn{1}{c}{$\Delta v_{\rm rad}$} &
\multicolumn{1}{l}{\phn$\sigma_{v_{\rm rad}}$} &
HIP & &
\multicolumn{1}{c}{$\Delta v_{\rm rad}$} &
\multicolumn{1}{c}{$\sigma_{v_{\rm rad}}$}
\\
no. & \multicolumn{1}{r}{\raisebox{1.5ex}[-1.5ex]{$n_{\rm obs}$}} &
\multicolumn{1}{c}{[\ms]} &
\multicolumn{1}{l}{[\ms]} &
no. & \multicolumn{1}{r}{\raisebox{1.5ex}[-1.5ex]{$n_{\rm obs}$}} &
\multicolumn{1}{c}{[\ms]} &
\multicolumn{1}{l}{[\ms]} &
no. & \multicolumn{1}{r}{\raisebox{1.5ex}[-1.5ex]{$n_{\rm obs}$}} &
\multicolumn{1}{c}{[\ms]} &
\multicolumn{1}{c}{[\ms]}
}
\startdata
   379                  &  5  &   7.0 &   36.3  & 47959                  &  3  &   4.3 &   42.8  &  84950                  &  6  &   5.0 &  145.2 \\
  2497                  &  6  &   7.1 &   29.9  & 50027                  &  4  &   6.7 &   11.6  &  85139                  &  7  &   9.0 &   57.8 \\
  6732                  &  7  &   5.9 &   30.8  & 50336                  &  3  &   5.4 &   68.4  &  85888                  &  6  &   6.3 &   11.8 \\
  9110\tablenotemark{a} &  4  &   4.9 &  137.1  & 53261                  &  4  &   4.7 &   80.2  &  88636                  &  6  &   5.9 &   30.3 \\
 11432                  &  8  &   8.0 &   18.7  & 53316                  &  4  &   6.2 &    9.6  &  88684                  &  6  &   5.6 &   11.1 \\
 13905                  &  7  &   6.1 &   23.0  & 53781                  &  4  &   5.1 &  184.4  &  88839                  &  7  &   6.1 &   27.9 \\
 15861                  &  5  &   5.1 &  370.9  & 55086                  &  4  &   5.6 &   24.9  &  90067                  &  7  &   5.1 &   40.8 \\
 19011                  &  8  &   7.3 &   23.3  & 55716                  &  4  &   4.3 &   67.5  &  91004                  &  6  &   7.2 &  250.5 \\
 19388                  &  8  &   6.5 &   14.3  & 58181                  &  5  &   6.9 &   13.4  &  92747                  &  7  &   7.9 &  442.4 \\
 22860                  &  7  &   7.5 &   12.1  & 59847 &  5  &   8.0 &    0.0\tablenotemark{b}  &  93026                  &  6  &\nodata& \nodata\\
 30457                  &  6  &   5.7 &   21.6  & 61420\tablenotemark{c} &  3  &\nodata& \nodata &  96459                  &  8  &   6.0 &    4.3 \\
 30720                  &  3  &   2.0 &    7.4  & 61571                  &  5  &   6.4 &   21.1  & 101986                  &  8  &   6.5 &   52.8 \\
 32814                  &  3  &   4.0 &   19.9  & 64078                  &  5  &   9.3 &   27.2  & 105497                  &  8  &   5.9 &   21.4 \\
 33914                  &  3  &   3.8 &    1.2  & 64823                  &  4  &   4.7 &   13.4  & 108691                  &  8  &   6.5 &   27.3 \\
 34033                  &  4  &   4.7 &   25.7  & 65323                  &  4  &   6.0 &   84.0  & 109023                  &  7  &   5.8 &   20.2 \\
 34387                  &  3  &   3.5 &   12.9  & 72210                  &  4  &   6.1 &   26.5  & 109068                  &  8  &   4.2 &   27.3 \\
 35907                  &  3  &   2.9 &   50.4  & 73133                  &  4  &   5.6 &   63.8  & 109602                  &  7  &   6.1 &   30.4 \\
 36388                  &  3  &   3.7 &   39.4  & 74239                  &  3  &   9.7 &   29.8  & 110986                  &  5  &   5.4 &   91.8 \\
 36616                  &  3  &   2.7 &   90.5  & 74732                  &  4  &   4.2 &   17.5  & 111944                  &  6  &   3.6 &   39.7 \\
 36848                  &  3  &   5.2 &   22.7  & 75730                  &  3  &   2.8 &   58.4  & 112067                  &  6  &   8.5 &   18.6 \\
 38253                  &  3  &   4.3 &  164.6  & 75944                  &  4  &   6.6 &   28.0  & 113084                  &  6  &   6.2 &   21.0 \\
 38375                  &  3  &   4.1 &    4.2  & 78132                  &  4  &   4.5 &    7.7  & 113562                  &  6  &   8.0 &  218.7 \\
 39079                  &  4  &   4.8 &   28.2  & 78442                  &  4  &   5.0 &   11.0  & 113622                  &  6  &   5.3 &   18.2 \\
 39177                  &  3  &   4.6 &   89.3  & 79195\tablenotemark{d} &  3  &   3.0 & 2179.9  & 113686                  &  6  &   7.7 &   35.7 \\
 41909                  &  3  &   2.9 &   13.0  & 79540                  &  7  &   6.0 &   31.9  & 113864                  &  7  &   5.3 &   22.7 \\
 43923                  &  3  &   3.4 &    4.6  & 80693                  &  7  &   5.2 &   64.2  & 114449                  &  6  &   5.9 &   14.8 \\
 44356                  &  4  &   7.1 &   46.0  & 81660                  &  6  &   4.7 &   14.4  & 117567                  &  6  &   6.6 &   61.4 \\
 46982                  &  4  &   7.9 &   18.6  & 83254                  &  7  &   6.4 &  100.5  & 117756                  &  5  &   6.5 &    8.8 \\
 47189                  &  3  &   4.3 &  136.5  & 84671                  &  6  &   5.0 &   45.2  \\
\enddata
\tablenotetext{a}{SB: \citealt{gordon46,irwin52}}
\tablenotetext{b}{observed velocity dispersion is 7.4\,\ms, smaller than the mean error of 8.0\,\ms}
\tablenotetext{c}{SB: \citealt{griffin92}}
\tablenotetext{d}{SB: \citealt{ginestet85}}
\tablecomments{Given are the number of observations for each star, the mean measurement
error of the radial velocities, and the intrinsic dispersion. The observed dispersion is
the quadratic sum of the last two columns. For stars with only three observations,
the measurement errors are somewhat optimistic and will get readjusted once
more observations are available for them.
For two of the stars it was not possible
to calculate the radial velocities, i.e.\ the fit of Doppler shifted template star 
spectrum plus iodine did not yield reasonable fits to the individual iodine star spectra.
However, this only happens for very large Doppler shifts.}
\end{deluxetable}

\begin{figure*}
\epsscale{1.5}
\plotone{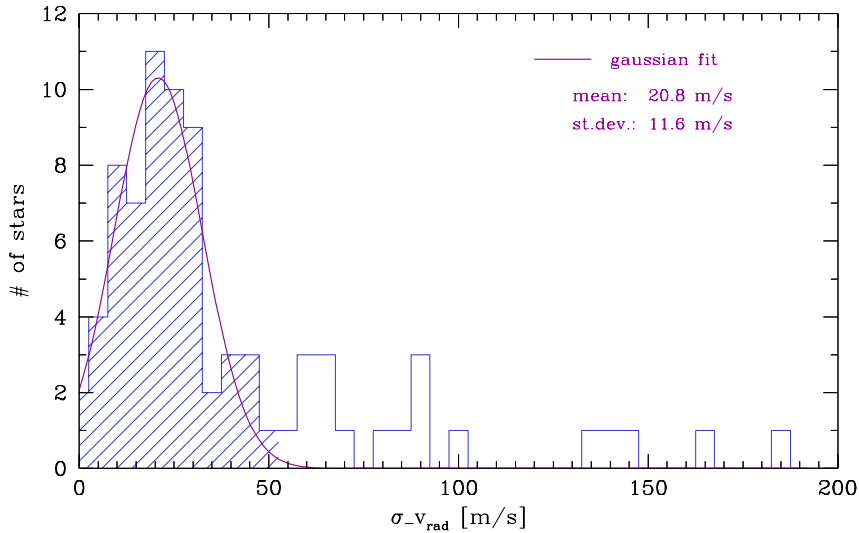}
\figcaption{\label{disp} Observed intrinsic radial velocity dispersions in our
proxy sample of 86 nearby \hipparcos\ K~giants as listed in Table~\ref{obstab} 
(histogram). Five~stars with
dispersion larger than 200\,\ms\ are not shown. The solid line is a gaussian
fit to the distribution of dispersions smaller than 50\,\ms\ only (hatched part
of the histogram) and reveals a maximum around 20\,\ms.
}
\end{figure*}

\begin{figure*}[t]
\epsscale{2}
\plotone{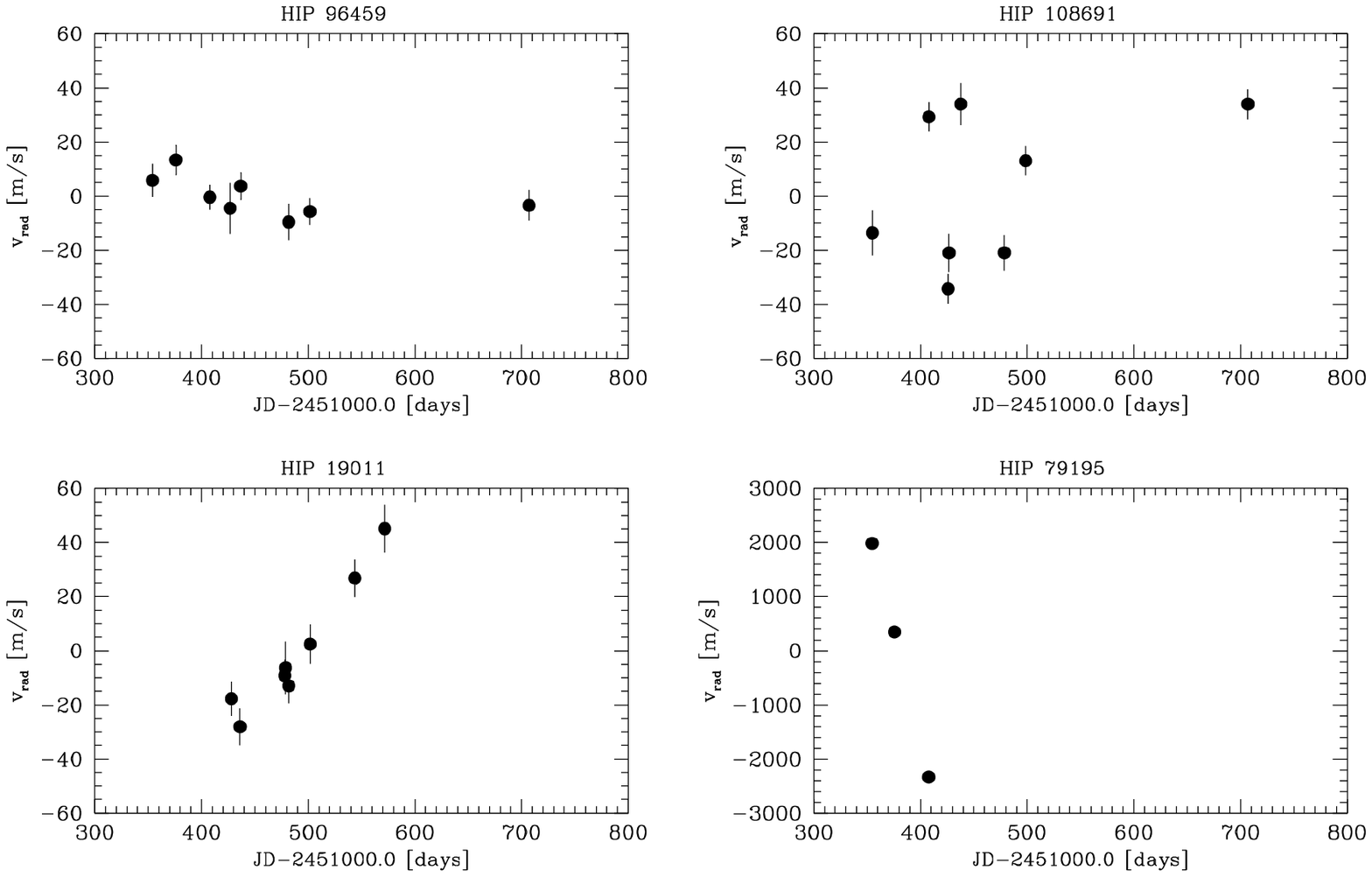}
\figcaption{\label{examp} Observed radial velocities (with arbitrary zero points)
versus time for four stars of the CAT proxy sample.
HIP~96459 is one of the stars with the smallest radial
velocity scatter in our sample; the observations are consistent with a
constant radial velocity within the errors. HIP~108691 is a more typical star;
the intrinsic velocity scatter is 23\,\ms, and the velocity pattern looks
random. In contrast to this, the radial velocities follow a trend for
HIP~19011, although its overall intrinsic velocity dispersion is even a little
lower than the one for HIP~108691. Finally, HIP~79195 is an example for a spectroscopic
binary which is easily identified. Note the different scale in the last panel,
which makes the error bars smaller than the plotting symbols.
}
\end{figure*}

In order to demonstrate the feasibility of our approach to effectively sort out
the double stars from a sample of candidate K~giant grid stars we started to
measure precise radial velocities for a number of stars in the nearby 
\hipparcos\ sample defined in the last section. 
The observations were carried out with the
0.6\,m Coud\'e Auxiliary Telescope (CAT) at Lick Observatory using the Hamilton
Echelle Spectrograph in conjunction with the Iodine Cell. The spectra were
reduced in the same way as described in \citet{butler96},
which has been proven to yield accuracies up to 3\,\ms. Exposure times are up
to half an hour for 6\,mag stars, which results in a typical signal-to-noise
ratio of 100 (not lower than about 80, in a few cases as high as 150 or even 200).

We started the monitoring in June~1999 and typically observed for five consecutive
nights every month, so that we are able to probe the intrinsic radial velocity
variations on timescales from days up to one year so far. Altogether there are
177~\hipparcos\ K~giants brighter than 6\,mag that passed our selection
process, and 139 of them are accessible from Lick Observatory ($-30\arcdeg \leq 
\delta \leq +68\arcdeg$). Out of these 139 K~giants we chose 86 to comprise our
proxy sample, including all stars brighter than 5.5\,mag. Between 5.5\,mag and
6\,mag we chose the ones which were most convenient to observe, i.e. according
to position on the sky which should not have introduced any biases.

For all stars in our proxy sample we have at least three (typically five or six) 
observations plus template so far.
Typical examples are shown in Fig.~\ref{examp}, and all resulting intrinsic
radial velocity dispersions are listed in Table~\ref{obstab}.

The first thing to note is that there are still a few spectroscopic binaries
present in our sample, which was not obvious from the data given in the
Hipparcos Catalogue. Most of them are known spectroscopic binaries, but one or
two could be new ones.
However, all of them could easily be identified with only two
or three observations. Discarding these spectroscopic binaries from our sample, the
distribution of intrinsic radial velocity variations shows a peak
around 20~\ms\ (Fig.~\ref{disp}). The distribution can be fitted by a gaussian
with a mean of 20.8\,\ms\ and a standard deviation of 11.6\,\ms.
This overall low level of radial velocity variability is remarkable, since it
is known from intensive monitoring of a few bright K~giants that they can
display radial velocity amplitudes of several 100~\ms.

\begin{figure}[t]
\epsscale{1}
\plotone{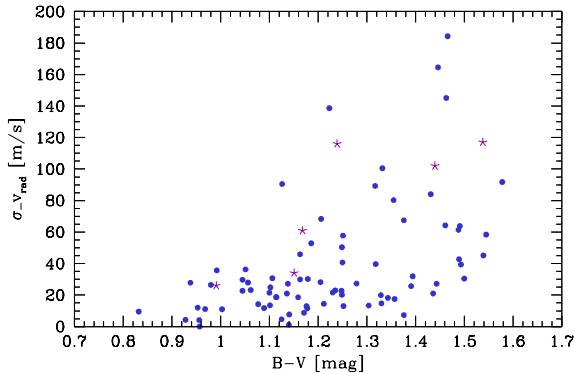}
\figcaption{\label{color} Dependence of the observed intrinsic radial velocity
variability on $B-V$ color, taken from the Hipparcos Catalogue,
for the stars in our proxy sample (filled circles). K~giants from
other investigations mentioned in the text are also shown
(asterisks). There is a trend visible in the sense that redder
K~giants show larger radial velocity dispersions.
}
\end{figure}

There is a trend with color in the observed radial velocity dispersions
in the sense that the redder K~giants show larger variations
(Fig.~\ref{color}). This is not too surprising, since these late
K~giants are located adjacent to the highly variable M~giants in the
Hertzsprung-Russell diagram. For the selection of the candidate grid
stars it might therefore be best to apply a certain color cutoff
around $B-V=1.2$\,mag.

\subsection{Comparison with other studies}

The first and best studied K~giant with known radial velocity variability is
probably Arcturus ($\alpha$~Boo).
It was shown to have several short-term periods of the
order of a few days by \citet{smith87} and \citet{hatzes94a}, 
as well as a long-term period of
233\,days with a radial velocity amplitude of 500\,\ms\ 
\citep{hatzes93}. Similarly, short- and long-term periods have
been derived for the K~giant $\beta$~Oph \citep{hatzes96}, 
and long-term periods of the order of
1.5--2~years for $\beta$~Gem, $\pi$~Her \citep{hatzes99}
and $\alpha$~Tau (with additional night-to-night
variations of 100\,\ms, \citealt{hatzes93}).
Moreover, \citet{walker89} observed a sample of five K~giants
and one K~supergiant (including Arcturus, $\beta$~Gem and $\alpha$~Tau), 
and {\it all} of the stars in this small sample showed
long-term radial velocity variations over the course of one year with amplitudes
between 30 and 300\,\ms, suggesting that these kind of variations are quite
common.

The most likely explanation for these variations are radial pulsations for the
short-term variations of a few days, and either non-radial pulsations 
or rotational modulation of surface features such as star spots for the
long-term variations. There are a few indications for the mechanism that is
responsible for the long-term variations. For example, \citet{hatzes94a} found
evidence for a mode-switching in $\alpha$~Boo, which is only possible if the
long-term variations are due to non-radial pulsations. However, $\alpha$~Boo
and possibly $\beta$~Oph \citep{hatzes94b} seem 
to be the only K~giants for which period changes have been observed so far.
On the other hand, \citet{walker89} found a correlation of the
long-term variations with chromospheric activity,
\citet{lambert87} observed variations in the \ion{He}{1} line which
were correlated with the radial velocity variations of $\alpha$~Boo, and 
\citet{walker92} found similar changes in the equivalent width of the
\ion{Ca}{1} line of $\gamma$~Cep. 
These spectral line variations are likely to be caused by the rotation of the
star, and the related periods would correspond directly to the rotation period. 
Since the radial velocity variations show the same periodicity, this would clearly
speak for rotational modulation of surface features as the mechanism producing
these variations.

In principle there is a third possible explanation for the radial velocity variations:
they could be caused by planetary companions. However, given the observed
correlations with variability of spectral features, mode switching or photometric
variability other explanations seem to be more likely, although planetary
companions could still be present at least in a few cases. They could probably
be detected with higher confidence if they were in rather elliptic orbits,
imprinting the well-known characteristic shapes in the radial velocity curves.
Another diagnostic tool which could help to detect the mechanism causing the
radial velocity dispersions are spectral line bisector variations,
which measure asymmetries in the spectral line profiles; they require rather
high resolution and signal-to-noise data. Bisector variations are expected for both
non-radial pulsations and rotational modulation, but not for planetary
companions \citep{hatzes96b}.

\citet{walker89} noted that all six K~giants in their sample
were photometrically variable. This might be an important difference to the
K~giants in our proxy sample, which were selected not be detected as
photometrically variable by \hipparcos. 

Our results show that it is possible to find K~giants with much
smaller variations, at least on timescales up to one year and in a sample
strongly biased against active, variable and multiple stars.
The observed radial velocity dispersions are not expected to become much
larger on longer timescales, since we already probed timescales of one year
which is close to the observed periods for other K~giants. 
For 18~stars we already have second year data, and for eight out of ten stars
with intrinsic velocity dispersions smaller than 50\,\ms\ these
dispersions changed by less than 5\,\ms, typically only 1--2\,\ms.
For one more star (HIP~105497)
the velocity dispersion changed by about 13\,\ms, but is
still very small, and only one (HIP~80693) of the stars which would have been 
classified as good grid candidate before shows considerably more variations now.

Although we cannot identify the mechanism producing the observed dispersions
with high confidence, it is unlikely that they are caused by planetary
companions, which would affect their grid star properties, in contrast to the
more likely explanations involving non-radial pulsations or rotational
modulation of surface features.
In any case, they are small enough to not severely affect the
identification of binary stars among K~giants, and we use the observed
distribution of radial velocity dispersions as input parameter for Monte-Carlo
simulations of radial velocity surveys designed to identify the
binaries in a K~giant sample.

\section{A Strategy for the Whole Grid}

After the selection of a candidate sample of K~giants that might be suitable
SIM grid stars it is essential for the astrometric stability of the grid 
to ensure the absence of stellar companions. 
While adaptive optics imaging and photometric monitoring could
help to further vet the candidate sample from unwanted perturbers, a precise radial
velocity survey of all grid stars is the most efficient way to identify a large fraction
of the multiple stars in the candidate sample.

The following sections deal with the design of such a radial velocity survey. 
Any such survey will require a huge amount of observing resources from the
ground, and it should be conducted as efficiently as possible.
A separate photometric survey might be needed to provide an input
sample for the radial velocity survey which is already cleaned from
variable stars, in case it should turn out that photometric
variability is correlated with intrinsic radial velocity variability
and photometric variability information is not already available for
the chosen kind of stars.

\subsection{Simulation of a Radial Velocity Survey}

We carried out simulations of a radial velocity survey of 3000~stars with an
assumed binary frequency of 50\% and determined the number of binaries that
would escape detection for various observing scenarios. We also counted
the number of single stars that would erroneously be rejected as grid stars, as
well as the number of missed binaries that would produce an
astrometric jitter larger than 1\,\muas\ in the \itsim\ solution.
For both single and double stars we
added intrinsic radial velocity variations with a gaussian distribution (mean
20\,\ms, dispersion 12\,\ms), in accordance with the results of our proxy
sample observations.

The orbit parameters for the binary stars were chosen to follow the observed
distributions for other samples. The logarithm of the period $P$ was generated
according to a gaussian distribution with a mean of 
$\overline{\log P[\mbox{days}]}=4.8$ and a dispersion of 
$\sigma_{\log P[\mbox{days}]}=2.3$ \citep{duquennoy91}. 
The eccentricities $e$ were distributed
according to $f(e)\,\mbox{d}e=2e\,\mbox{d}e$, as indicated in the same sample
from \citet{duquennoy91}. However, for
systems with periods less than 100~days the eccentricities were set to zero,
since \citet{boffin93} found a circulization period of about 
100\,days in a sample of
213~spectroscopic binaries with late-type giant primaries. Boffin et al.\ also
analyzed the distribution of secondary masses in their sample and found no
significant deviation from a uniform distribution. Hence, we assumed a fixed
mass of 1\,\msun\ for the K~giant primary in our simulated sample and
distributed the secondary masses uniformly in the mass range between 
0.08\,\msun\ and 1.0\,\msun.
Random orientation of the orbits in space leads to a uniform distribution of 
periastron lengths $\omega$ and a distribution of the inclination angle $i$ 
according to $f(i)\,\mbox{d}i=\sin i\,\mbox{d}i$. 
Face-on orbits with small inclinations are thus much less common
than edge-on orbits, which is favorable for the detection of radial velocity
signals.

The parameters that we varied include the radial velocity precision (from 0\,\ms\ to
200\,\ms), the number of observations for each star (between 2 and 10), and
the duration of the survey (up to 8~years). Furthermore, we adjusted the
threshold of the reduced $\chi^2$ (for a constant model radial velocity)
above which a system is flagged as suspected binary; this corresponds to 
constraining the tolerated number of rejected true single stars and also 
roughly to the total number of systems that have to be observed.

For the models with two observations per star, the spacing between them
corresponds to the survey duration. For the other models the additional
observations were distributed randomly over the survey duration.

For the calculation of the astrometric jitter in the \itsim\ solution it was
assumed that each grid star will be observed twice during each of altogether
23~grid campaigns, spaced ten days apart. The interval between two
subsequent grid campaigns was set to 40~days. A linear least squares fit to the
observed positions was performed, which allowed
part of the orbital motion to be absorbed as proper motion. The remaining
scatter in the positions, corresponding
roughly to half the value of the largest deviation of the star from a line
connecting its position at the beginning and at the end of the mission,
was used as a measure for the astrometric jitter.

For each parameter set we performed 10\,000 simulations and obtained the mean 
percentages of truly single stars, unidentified binaries and unidentified
binaries that would produce an astrometric jitter larger than 1\,\muas\
in the final composition of the grid. We also determined dispersions around
these mean values for various realizations of samples with the same input parameters.

The parameter values for our reference model are as follows:
\begin{list}{$-$}{\leftmargin0.7cm\topsep0.2ex\itemsep-0.8ex}
\item duration of the survey: 5~years
\item measurement accuracy: 20\,\ms
\item number of observations per star: 2
\item tolerated fraction of rejected single stars: 32.5\%
\end{list}
This results in a binary fraction of 32.2$\pm$0.9\% in the final grid sample,
with 3.6$\pm$0.5\% of all grid stars displaying astrometric jitter in excess of
1\,\muas. It would require to observe two times as many stars as ultimately 
needed for the grid.

\begin{figure}
\plotone{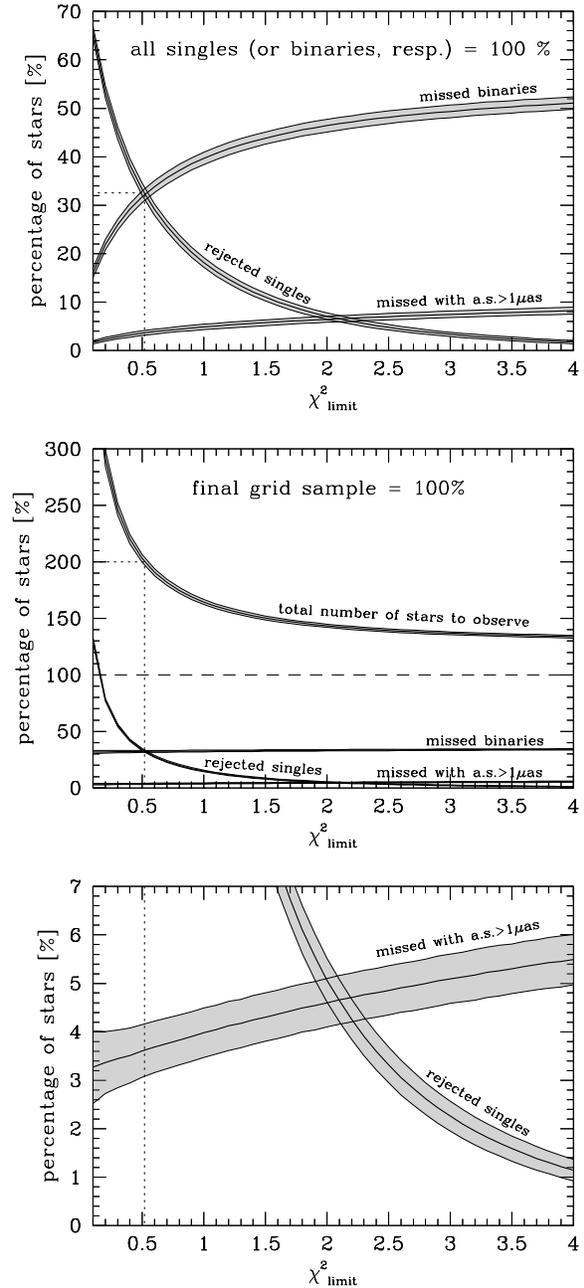}
\figcaption{\label{cutdisp} Fractions of single and binary stars that would be
missed in a radial velocity survey as a function of the threshold in the 
computed $\chi^{2}$. Values in the upper panel are normalized to the respective
number of single and binary stars in the input sample, whereas in the other two
panels they are normalized to the final composition of the grid.
The dotted lines indicate the value of $\chi^{2}{}_{\rm limit}$ 
that corresponds to observing twice as many stars as needed for the grid.}
\end{figure}

\subsection{Discussion of the Simulation Results}

The results of our simulations are presented in Figs.~\ref{cutdisp}, \ref{simu}
and \ref{hist}. Fig.~\ref{cutdisp} illustrates the fractions of stars that
will be erroneously rejected as single stars or missed as binary stars, 
respectively, as a function of the threshold in the computed $\chi^2$. 
The top panel shows the results normalized to the total number of single and 
binary stars in the input sample. If one is willing to tolerate a success rate
in identifying true single stars of only 50\%, this results in a lower fraction
of missed binaries of 23.4\% as compared to 32.1\% in our reference model, which only 
allows a fraction of 32.5\% of the true single stars to be rejected as grid stars.

However, such a high rejection rate of single stars would vastly increase the
number of stars in the input sample for the radial velocity survey (see middle
panel of Fig.~\ref{cutdisp}).
In our reference model, we required that no more than twice as many stars have
to be observed than finally needed for the grid, resulting in a tolerated
rejection rate of single stars of 32.5\% (and a $\chi^2$ of 0.524 in the
reference model, indicated by the dotted lines in Fig.~\ref{cutdisp}).
For a tolerated rejection rate of 50\%, the radial velocities of about three 
times as many stars as finally needed for the grid would have to be analyzed.
Besides the problems of finding so many suitable grid candidate stars, this
dramatically increases the required observing resources, while only marginally
improving the results: though the success rate in identifying binaries improved
by about 10\%, the fraction of unidentified binaries present in the grid
will still be 31.9\%, as compared to 32.2\% in the reference model, and
the number of unidentified binaries with astrometric signatures larger than
1\,\muas\ decreases only from 3.6\% to 3.4\%. The reason for these very small
improvements is the increase in the number of stars in the input sample, which
requires a larger success rate for the identification of binaries in order to
achieve the same binary contamination fraction in the final grid. These effects
almost cancel out, leading to only very small improvements for the final grid.

However, the bottom panel of Fig.~\ref{cutdisp} (which is a closeup of the lowest
part of the middle panel) shows that there is still some merit in observing a 
larger number of stars. Lowering the tolerated rejection rate of single stars
to only 2\% would result in a 5.4\% contamination of the grid with
binaries that would produce an astrometric jitter larger than 1\,\muas, as
compared to 3.6\% for the reference model. The gain in the number of stars that
would have to be observed is not so large as to justify this degradation
(especially since every effort should be made from the ground to define a grid
as clean as possible, see below): still 1.3 times as many stars as needed for 
the grid would have to be observed.
Observing two times as many stars as finally needed for the grid is also a
reasonable number considering the fact that, given an assumed overall binary 
frequency of 50\%, there will be just as many true single stars in the input sample 
as needed for the grid, and therefore we adopted the corresponding value of 
32.5\% for the tolerated rejection rate as our reference value.

\subsection{Fine-tuning the parameters}

\begin{figure}
\plotone{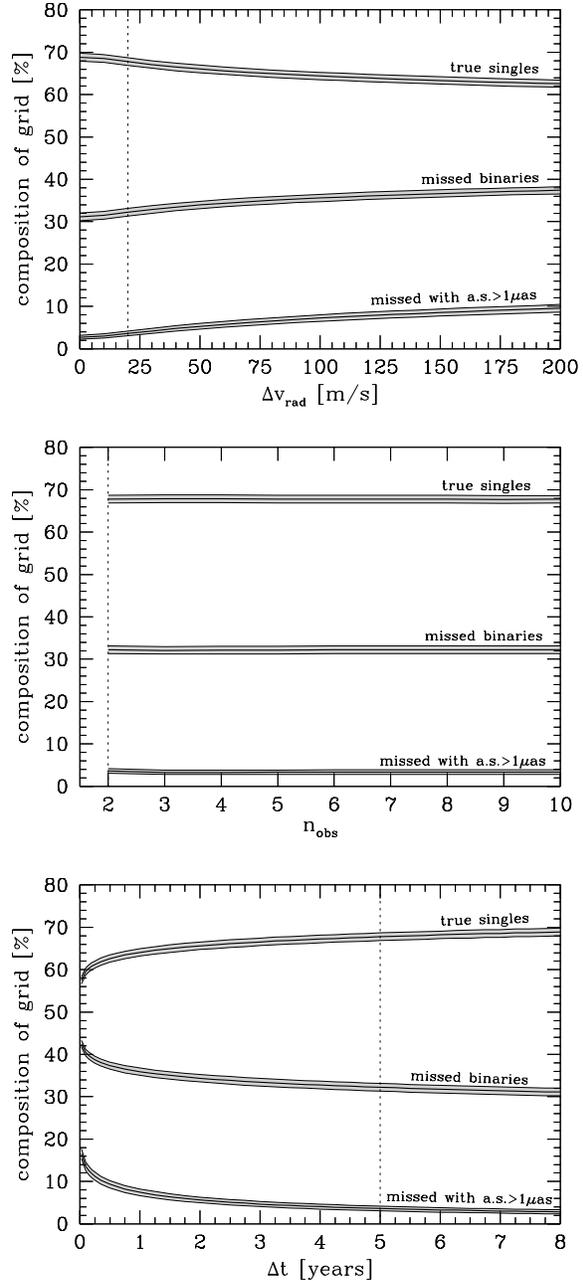}
\figcaption{\label{simu} Influence of radial velocity accuracy (upper panel),
number of observations (middle panel) and survey duration (lower panel) on the
final composition of the grid according to our simulations. The dotted lines
indicate the values of our reference model; see text for details.}
\end{figure}

Figure~\ref{simu} shows the influence of some basic parameters of the radial
velocity survey (accuracy of the radial velocities, the number of observations
and the duration of the survey) on the final composition of the grid.
As expected, the accuracy of the radial 
velocities has a mild influence on the
ability to reliably identify binary stars. In the theoretical case of
infinitely accurate radial velocities one still would end up with a 2.7\%
contamination in the grid by binary stars with astrometric jitter larger than
1\,\muas, while an accuracy of 200\,\ms\ would lead to a contamination level
with these stars of 9.6\%. The reason why there are still a number of
unwanted binaries left in the grid even in the case of infinitely accurate
radial velocities is the intrinsic radial velocity scatter for K~giants of
about 20\,\ms, as discussed in Sect.~\ref{proxy}. It does not seem very
meaningful to measure radial velocities which are significantly more precise
than the expected intrinsic scatter, and our reference value of 20\,\ms\ for the
precision of the radial velocity measurements reflects this situation. On the
other hand, 20\,\ms\ is exactly the accuracy which one could achieve with Keck
for a 12\,mag K~giant in a 2~minute exposure, already comparable to the 
readout time. At a 3m class telescope, one could probably achieve 50\,\ms\ in a
3~minute expsosure, but this would already result in a higher contamination
fraction in the final grid of 5.3\%. So 20\,\ms\ seems to be a good compromise,
resulting in a 3.6\% contamination of the grid with binary stars producing an
astrometric jitter larger than 1\,\muas.

\begin{figure*}[t]
\epsscale{1.8}
\plotone{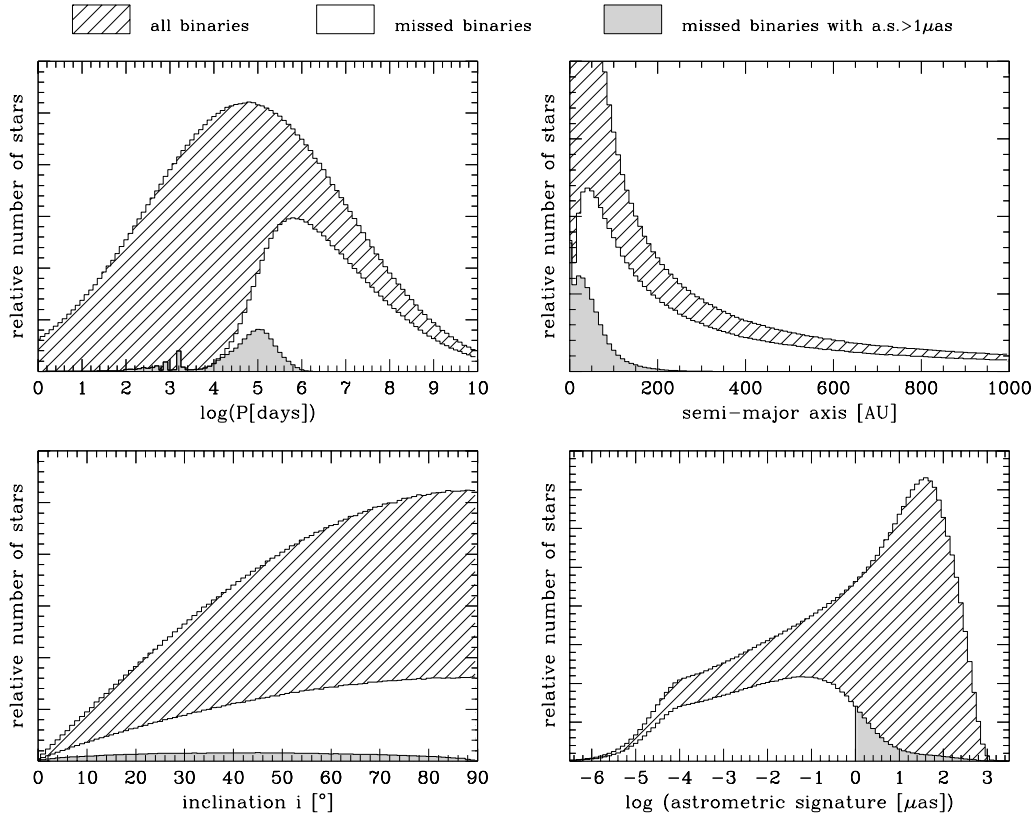}
\figcaption{\label{hist} Distribution of periods, semi-major axis, inclinations
and astrometric signatures as seen by \itsim\ in our simulated binary input
sample (hatched), among the sample of binaries that would be missed in our reference
radial velocity survey (white), and among the subset of the latter with
astrometric signatures larger than 1\,\muas\ (grey). The peaks in the period
distribution with periods of 5~years and less stem from observing
the star during the same phase of the orbit; they could be eliminated from the
final grid by scheduling one more observation for all stars. The lower right
plot nicely demonstrates the power of a radial velocity survey in identifying
the vast majority of stars that would cause major problems for the \itsim\ grid.}
\end{figure*}

The most surprising parameter in our reference model might be the low number of
only two observations per star. Indeed, at least if the survey is conducted with
sufficient precision, virtually every binary star up to a certain period can be
identified already by the change in radial velocity over a five year interval 
(see also Fig.~\ref{hist}). Additional observations in between these two 
observations only help to identify the very few binaries with periods which are
integer fractions of the interval between observations, so that in the scenario 
with three observations per star the contamination rate of the grid with
binaries with astrometric signatures larger than 1\,\muas\ is 3.3\% as compared 
to 3.6\% in the reference model. Adding even
more observation does not decrease the contamination fraction any further; it
basically remains constant for three observations onward for the models with a
radial velocity accuracy of 20\,\ms. For a degraded radial velocity accuracy, 
it takes more observations per star to achieve the lowest possible
contamination level,
though it will always be higher than the one for surveys with better precisions.
In other words, it is not possible to make up for lower precision radial
velocities with scheduling more observations per star. For example, with a
radial velocity precision of 50\,\ms\ the asymptotic contamination level of
4.7\% is reached with four observations per star, while with a radial velocity
precision of 200\,\ms\ a level of 7.9\% is only reached after
eight observations per star.

Finally, the longer the interval between the first and the last observation,
the higher the sensitivity towards the identification of longer period
binaries will be. An increase in the survey duration from five years as in our
reference model to eight years will decrease the contamination level of the final
grid with binaries producing astrometric jitter larger than 1\,\muas\ from
3.6\% to 2.8\%. A 20~year baseline for the radial velocities even would
decrease the fraction down to 1.9\%. This is quite remarkable, all the more
since it is achievable without any additional amount of observing resources. 
However, the value of five years which we chose for our reference model
reflects the maximum time that we realistically could assume to have left
before the launch of \itsim. 
On the other hand, scheduling the second observations only two years after the
first observations would result in a 5.6\% contamination fraction with binaries
producing astrometric jitter in excess of 1\,\muas.
Our simulation results underline the importance of a timespan as
large as possible in between individual radial velocity observations in order
to achieve the lowest possible contamination fraction of the grid with unwanted
stars. 

\subsection{Properties of missed binary stars}
It is very important to know not only how many binaries will be missed in a
radial velocity survey and contaminate the grid, but also the characteristics
of these multiple systems. We plotted the distributions of periods, semi-major
axis, inclinations and astrometric jitter as seen by \itsim\ in our generated
input sample as well as in the sample of binary stars that escape detection in
the radial velocity survey (Fig.~\ref{hist}). Those binaries left in the grid
that will produce the largest astrometric jitter are also marked. 

The first thing to note is that basically all binaries with periods less than
10\,000~days can be identified, except for a few systems with periods which are
integer fractions of the interval between observations (Fig.~\ref{hist}, upper left).
For systems with periods greater than 1\,million days (2700~years) the success 
rate in identifying the stars as binaries decreases drastically. 
Fortunately these are the systems that will not impose a major problem for the 
\itsim\ grid; their astrometric jitter will be small, since the periods are so 
large compared to the \itsim\ mission duration that no orbital curvature will
be detected. 

The systems that would produce a measurable astrometric jitter if included in 
the grid are those with periods around 100\,000~days. 
Corresponding semi-major axis are mostly less than 100\,AU, typically even less 
than 50\,AU (Fig.~\ref{hist}, upper right). Any additional observing
program aimed at further reducing the fraction of binary stars in the final
grid should be designed to be sensitive to these kind of systems. In the
optical, the magnitude difference between a K~giant and an M~dwarf could be as
high as 8--12\,mag, which would make direct detections rather
difficult. Infrared adaptive optics imaging might offer a favorable advantage.

The Tycho-3 project which is now starting in Copenhagen could also be of use
for the selection of \itsim\ grid stars. While the variability data in Tycho-2
are rather incomplete especially at the faint end relevant for the \itsim\
grid, this situation should vastly improve in Tycho-3. Tycho-3 will also
supply duplicity data which could help to eliminate wide binaries from the
grid; however, the large number of shorter period binaries, causing greater
astrometric signatures when included in the grid, will still escape detection in
the Tycho-3 reductions due to the large distance of the grid stars.

For a number of systems with projected separations in the range 
$\approx 0.01-1$\arcsec\ and small magnitude differences \itsim\ would 
detect double star fringes. However, this is not a big concern since those
stars could be eliminated from the grid after only one or two \itsim\ observations.

The distribution of astrometric jitter as seen by \itsim\ in the binary input
sample as compared to the sample of 
unidentified binaries (Fig.~\ref{hist}, lower right) illustrates nicely the
power of a radial velocity survey in identifying preferrably those stars which
would cause the largest astrometric jitter. Without such a survey, the \itsim\
grid would encounter serious problems.

\section{Discussion}
Cleaning up a pre-selected
sample of candidate grid stars from binaries with a radial velocity survey 
requires a huge amount of observing resources; estimates are 100~nights at a
10\,m telescope or 500~nights at a 3\,m telescope for 1500~suitable grid stars.
If we knew for sure that the resulting contamination fraction in the grid would
not be larger than that derived in our Monte-Carlo simulations in the last
section, it probably would not be justified to perform the radial velocity
survey with such a high accuracy as assumed in our reference model.

However, there are quite a number of uncertainties involved that could increase
the number of unusable grid stars.

First, we do not know whether the true binary frequency and
the distribution of orbital parameters in a grid candidate input sample match
those that we used for our Monte Carlo simulation. They were derived for a
sample of main sequence field stars and could be quite different for K~giants
in general and for Halo K~giants in particular, although there are no
indications for this so far \citep{abt83,harris83,boffin93,latham01}.
Shifting the maximum in the period distribution to only slightly larger
values however could have meaurable effects for the stability of the grid
(cf.~Fig.~\ref{hist}, upper left).

Second, we are not able to identify planetary companions to such a large number
of stars. Jupiter-mass companions orbiting within 1~AU will not cause a
measurable astrometric jitter for \itsim\ if the grid objects are
located at large enough distances, i.e.\ larger than about 2\,kpc. However, if
giant Jupiters in orbits of the order of 5~AU are common this will add
to the overall number of unusable grid stars.

Third, stars can be unsuitable as grid stars for other reasons than duplicity.
Phenomena that could produce photocenter shifts include
starspots, planetary transits and astrometric microlensing. The angular
diameter of a K~giant at 2\,kpc is of the order of 50\,\muas. Thus, in
order to produce a measurable astrometric jitter for \itsim, the photocenter
shift would have to be of the order of 10--20\,\% of the stellar radius. 
Most starspots are probably not
large and cool enough to produce such a giant shift, but there might be a few
rare cases where this appears to be possible.
\citet{strassmeier99} determined the size of the largest
starspot known to date on an active and variable K~giant; it covers about 
11\,\% of the entire stellar surface. While such spots are probably not very
common, the chances for facing this problem in the \itsim\ grid could be
minimized by a pre-selection of candidate grid stars in the input sample
against variable and active stars.

Planetary transits are not a big concern either. The radius of a giant
planet is about hundred times smaller than that of a K~giant, and even if the
planet completely occulted part of the stellar surface the induced positional
offset in the measured photocenter would be too small to be recognized by
\itsim.

The case might be different for astrometric microlensing. While photometric
microlensing is only detectable when the source and the lens are perfectly
aligned, the optical depth is much larger for astrometric microlensing which
aims at detecting directly the deflection of starlight as it passes close to
another body. \citet{miralda96} estimated that several
events could be detected with an astrometric precision of 10\,\muas\ by
monitoring many thousand stars over several years.

Finally, there might be additional phenomena causing astrometric
jitter that we have not thought of or do not know of yet. Since all these could
add to the contamination fraction in the \itsim\ grid, it is crucial for
the stability of the grid to identify as many problematic stars as possible
before the mission, from the ground. Any stars that are left in the grid that
are not as astrometrically stable as required could affect the observations
where these stars have been used as grid stars. Besides affecting individual
observations, a larger number of grid star observations have to be scheduled
along with each science observation for a higher contamination fraction in the
grid, taking precious \itsim\ observing time. Using \itsim\ itself for culling
unsuitable grid stars is clearly much more expensive than doing a thorough job
from the ground.

\section{Summary}

We have argued for K~giants as the best type of stars for use with the \itsim\
grid. Due to their high intrinsic luminosities K~giants can be located at much
larger distances than any other type of star. K~giants around 12\,mag would 
be located at about 2\,kpc for solar metallicity and up to 5\,kpc if they were
metal-weak, thus reducing any kind of astrometric jitter that could stem from
stellar and planetary companions as well as from other unknown sources.

We have shown that radial velocities can be measured
precisely enough to identify problematic companions to K~giants.
The intrinsic radial velocity scatter is of the order of 20\,\ms\ for
most of the stars in a proxy sample of \hipparcos\ K~giants which is highly
biased against multiple and variable stars. 
A few spectroscopic binaries that were included in our sample due to lack of
information in the Hipparcos Catalogue were easily identifiable as such with only
two or three observations. We are in the process of extending
our precise radial velocity measurements to a statistically unbiased sample of
K~giants. 

We simulated the possible design of a radial velocity survey of several
thousand stars. We find that the measurement precision and the timespan
over which the radial velocities are measured are of great importance, whereas
the number of observations has only marginal effects on the result. 
Our preferred scenario --- two radial velocities measured five years apart
with a precision of 20\,\ms\ for every star --- would result in a 32\% 
contamination of the final grid with binary stars, but only 3.6\% of the stars
in the grid would produce astrometric jitter larger than 1\,\muas.

Any such radial velocity survey would require large amounts of observing
resources. However, since the astrometric stability of the grid is a crucial
element for the success of the Space Interferometry Mission in its entirety,
any ground-based efforts ensuring the required astrometric quality of the grid 
seem to be justified.

\acknowledgements
Our sincere thanks go to Geoff Marcy and Paul Butler, who kindly let us use
their setup and reduction software, 
without which the precise radial velocity measurements of
this study would not have been possible. 
We would like to thank the referee, Sean Urban, for his comments which helped
to improve the manuscript.
S.F.\ and A.Q.\ gratefully acknowledge
support from NASA's SIM Preparatory Science Program.

\appendix
\section{Selection Criteria}
The criteria which we used to select the Hipparcos proxy sample and the Tycho-2
sample are described in Tables~\ref{hip_table} and \ref{tyc2_table},
respectively. They were designed to ensure that every possible star with low
astrometric quality or indications for duplicity or variability was rejected.
They are rather strict, and it is likely that a number of suitable stars also got
rejected. A number of the criteria are very similar, and many stars are
rejected by more than one criterion.
Most of the criteria listed refer to flags or parameters as given directly in
the Hipparcos, TRC, Tycho-1 and Tycho-2 catalogs, respectively.
However, in order to sort out possible astrometric binaries among our sample we
performed a comparison of the proper motions
in the Hipparcos and ACT/TRC catalogs, testing
whether the `instantaneous' Hipparcos proper motion (derived from
measurements collected over a timespan of 3~years) is consistent with the
proper motion determinations based on a larger epoch difference
($\approx$\,80\,years for ACT/TRC). Significant differences might reflect
orbital motions due to companions. We followed the approach by Wielen et al.\
(1999) who defined a test parameter $F$ which is basically the ratio of the
proper motion difference and its expected error in both directions of the
error ellipsoid. A star is classified as `single-star candidate' when $F <
2.49$, corresponding to a 2$\sigma$~criterion.
This lead to the criteria~14--17 listed in Table~\ref{hip_table}.

\begin{deluxetable}{cllc}
\tabletypesize{\footnotesize}
\tablewidth{0pt}
\tablecolumns{4}
\tablecaption{\label{hip_table} Selection criteria applied to Hipparcos stars}
\tablehead{\multicolumn{4}{c}{astrometry}}
\startdata
1 & H29 &   [percentage of rejected data] & 0\% \\
2 & H30 &   [goodness-of-fit statistic]   & $< 3$ \\
3 & q\_pm & [TRC quality flag for proper motion] & $\leq 2$ \\
\cutinhead{variability}
4 & H6 &  [coarse variability flag] & $\sqcup$ \\
5 & H46 & [scatter of Hp observations] & $<0.1$\,mag \\
6 & H52 & [type of variability] & $\sqcup$ \\
7 & T47 & [previously known or suspected as variable] & $\sqcup$ \\
8 & T48 & [variability of the Tycho measurements] & $\sqcup$ \\
\cutinhead{duplicity}
9 & H2 &  [proximity flag] & $\sqcup$ \\
10 & H55 & [CCDM identifier] & $\sqcup$ \\
11 & H59 & [double and multiple systems annex flag] & $\sqcup$ \\
12 & H61 & [suspected non-single] & $\sqcup$ \\
13 & ACTflg & [proper motion difference TRC-ACT] & $\sqcup$ \\
14 & $\Delta\mu_{\mbox{\tiny TRC,HIP}}$ & [total proper motion difference TRC-HIP] & $< 10$\,mas/a \\
15 & $F_{\mbox{\tiny TRC,HIP}}$ & [significance of $\Delta\mu_{\mbox{\tiny TRC,HIP}}$] & $< 2.49$ \\
16 & $\Delta\mu_{\mbox{\tiny ACT,HIP}}$ & [total proper motion difference ACT-HIP] & $< 10$\,mas/a \\
17 & $F_{\mbox{\tiny ACT,HIP}}$ & [significance of $\Delta\mu_{\mbox{\tiny ACT,HIP}}$] & $< 2.49$ \\[0.2ex]
\enddata
\end{deluxetable}

\begin{deluxetable}{cllc}
\tabletypesize{\footnotesize}
\tablewidth{0pt}
\tablecolumns{4}
\tablecaption{\label{tyc2_table} Selection criteria applied to Tycho-2 stars.}
\tablehead{\multicolumn{4}{c}{astrometry}}
\startdata
%\cutinhead{astrometry}
1 & g\_mRA & [goodness of fit for mean RA] & $\leq 2$ \\
2 & g\_mDE & [goodness of fit for mean Dec] & $\leq 2$ \\
3 & g\_pmRA & [goodness of fit for pmRA] & $\leq 2$ \\
4 & g\_pmDE & [goodness of fit for pmDE] & $\leq 2$ \\[0.2ex]
\cutinhead{photometry}
5 & $\sqrt{e\_\bt^{2}+e\_\vt^2}$ &  [combined standard error in \bt--\vt] & $\leq 0.3$\,mag \\[0.2ex]
\cutinhead{duplicity}
6 & pflag & [mean position flag] & $\sqcup$ \\
7 & prox & [proximity indicator] & 999 \\
8 & posflg & [type of Tycho-2 solution] & $\sqcup$ \\
\enddata
\end{deluxetable}

\end{document}